\documentclass[seceq]{ptptex}

\usepackage{graphicx}

\usepackage{wrapft}

\pubinfo{Vol.~120, No.~5, November 2008}

\title{
Formation and Radiative Feedback of First Objects and First Galaxies
}

\author{
Masayuki Umemura$^{1,}$\footnote{E-mail: umemura@ccs.tsukuba.ac.jp}, 
Hajime Susa$^{2,}$, Kenji Hasegawa$^{1,}$, Tamon Suwa$^{3,}$, and Benoit Semelin$^{4,5}$
}

\inst{
$^1$ Center for Computational Sciences, University of Tsukuba, 
Tsukuba 305-8577, Japan \\
$^2$ Department of Physics, Konan University, Okamoto, Kobe, Hyogo 658-8501, Japan\\
$^3$ Fujitsu Limited, Kawasaki, Kanagawa 211-8588, Japan\\
$^4$ LERMA, Observatoire de Paris, CNRS, 61 Av. de l' Observatoire, 75014
Paris, France \\
$^5$ Universit$\acute{e}$ Pierre et Marie Curie, 4 place Jussieu, 75005
Paris, France
}

\recdate{
; revised }

\abst{
First, the formation of first objects driven by dark matter
is revisited by high-resolution hydrodynamic simulations.
It is revealed that dark matter haloes of $\sim 10^4M_\odot$
can produce first luminous objects with the aid of dark matter cusps. 
Therefore, the mass of first objects is smaller by roughly
two orders of magnitude than in the previous prediction.
This implies that the number of Pop III stars formed in the early universe 
could be significantly larger than hitherto thought.
Secondly, the feedback by photo-ionization and photo-dissociation photons 
in the first objects is explored with radiation hydrodynamic simulations,
and it is demonstrated that multiple stars can form in a $10^5M_\odot$ halo.
Thirdly, the fragmentation of an accretion disk around a primordial protostar 
is explored with photo-dissociation feedback. As a result, it is found that
the photo-dissociation can reduce the mass accretion rate onto
protostars. Also, protostars as small as 0.8$M_\odot$ may be ejected
and evolve with keeping their mass, which might be detected as
``real first stars'' in the Galactic halo.
Finally, state-of-the-art radiation hydrodynamic
simulations are performed to investigate the internal ionization of 
first galaxies and the escape of ionizing photons. 
We find that UV feedback by forming massive stars enhances the escape fraction even 
in a halo as massive as $> 6\times 10^9M_\odot$, while it reduces the star formation
rate significantly. This may have a momentous impact on the cosmic reionization.
}

\begin{document}
\maketitle

\section{Introduction}

The first generation objects and galaxies in the early universe are 
of great significance as the progenitors of present-day galaxies,
the sources for cosmic reionization, the origin of heavy elements, 
and the generators of seed black holes for supermassive black holes inhabiting galactic centers. 

The formation of first generation objects 
has been explored by many authors. Originally, the minimum halo mass that can undergo 
gravitational instability was estimated to be from  $5 \times 10^4M_\odot$ 
at redshift $z\sim100$ to $5\times 10^7M_\odot$ at $z\sim 10$ 
from analytic arguments \cite{Tegmark97}, where $M_\odot$ is the solar mass. 
Later on, three-dimensional hydrodynamic simulations \cite{Fuller00} 
showed the minimum mass to be from $10^5M_\odot$ 
at redshift $z\sim100$ to $10^6M_\odot$ at $z\sim 10$. 
Then, full cosmological hydrodynamic simulations have been performed to show 
that the minimum mass is a weak function of redshift, 
which is  $\sim 7\times 10^5M_\odot$ in halo mass and 
$\sim 10^5M_\odot$ in baryonic mass\cite{Yoshida03}. 
As for the secondary collapse of first objects under the influence
of preforming stars, the radiative feedback through
photo-ionization of hydrogen, photo-dissociation of hydrogen molecules, and their combination
are important. 
This issue has been explored by the radiation hydrodynamics, where
hydrodynamic simulations are coupled with radiative transfer calculations
\cite{SU06,Hasegawa09,Susa09}.

Also, in the last decade, the mass of Population III (Pop III) stars
has been studied extensively.
It has been shown that first stars are likely to be 
as massive as $100-1000M_\odot$ \cite{BCL99,Abel00,Abel02,Yoshida08}.
On the other hand, if the collapse of a first object into a flat disk is considered,
first stars may form in a bimodal fashion with peaks of $\sim 1M_\odot$ and 
several $100M_\odot$ \cite{NU01}. 
Besides, the mass of massive Pop III stars could be reduced to $20-40M_\odot$
by cosmic variance\cite{ON07}, external feedback\cite{Susa09}, 
or internal feedback\cite{Hosokawa11}.
These works focused on the runaway collapse phase of first objects.
Very recently it has been shown that in the accretion phase 
the disk around a first protostar can
fragment into smaller pieces, eventually allowing the formation of
less massive stars down to $\sim 1M_\odot$ or 
subsolar value\cite{Stacy10,Clark11,Greif11}. 
Even in such a fragmented primordial disk, stars that form earlier can
make a significant impact through radiation hydrodynamic feedback
on subsequent star formation.

The first generation and second generation stars are assembled
into first galaxies. Ultraviolet (UV) radiation from massive stars
in first galaxies can play an important role not only 
on the following star formation history but also as the source for
cosmic reionization. 
The escape of ionizing photons from first galaxies is regulated by
the ionization of interstellar medium. This issue
should be also investigated with radiation hydrodynamics.

In this paper, we present recent high-resolution hydrodynamic simulations
on first objects and radiation hydrodynamic simulations on the radiative feedback in these objects.
Also, the fragmentation of an accretion disk around a first protostar 
is explored with photo-dissociation feedback, and radiation hydrodynamic
simulations are performed to investigate the internal ionization of 
first galaxies and the escape of ionizing photons. 
In \S 2, a dedicated simulator for radiation hydrodynamics, which has been
developed for the present purpose, is described. 
In \S 3, the collapse of first objects driven by dark matter cusps is explored.
The radiation hydrodynamic feedback there is studied in In \S 4.
In \S 5, the fragmentation of a first protostellar disk and 
radiative feedback by photo-dissociation photons is explored with
radiation hydrodynamic simulations.
In \S 6, the photo-ionization of first galaxies and the escape fraction
of ionizing photons are analyzed based on radiation hydrodynamic simulations.
\S 7 is devoted to the summary.

\section{{\it FIRST} Project}

\begin{figure}[t]
\begin{center}
\includegraphics[width=12.5cm]{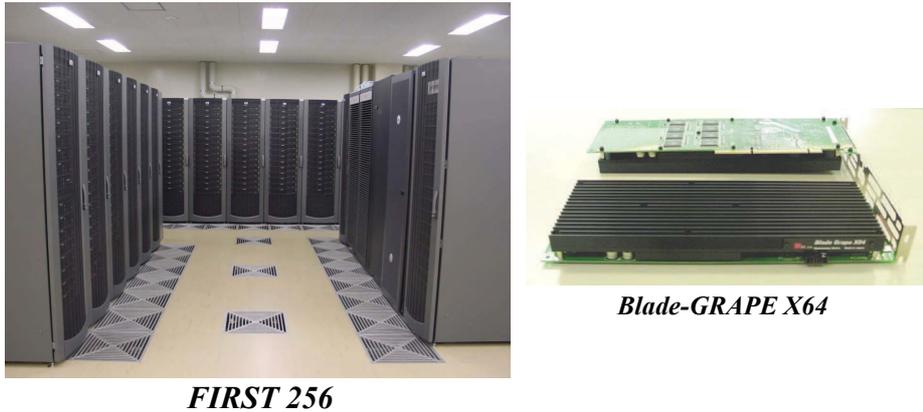}
\caption{The {\it FIRST} Simulator composed of 256 nodes ({\it left}) and an accelerator,
Blade-GRAPE, for gravity calculations, which is embedded in each node ({\it right}). }
\end{center}
\label{FIRST}
\end{figure}

In order to explore extensively the hydrodynamic and radiation hydrodynamic
processes during the formation of first objects and first galaxies,
we have built up a dedicated simulator called {\it FIRST}, on the basis of
the {\it FIRST} project \cite{Umemura08b}.
  
The {\it FIRST} project was initiated by a Specially Promoted Research in Grants-in-Aid for Scientific 
Research by MEXT over four years (2004-2007) with a budget of JPY 329.5 million (US \$4.1 million),
and has been continued with a Grant-in-Aid for Scientific Research (S) (2008-2012) by JSPS with a budget of JPY 73.1 
million (US \$0.9 million). In this project, we constructed a new hybrid simulator {\it FIRST} (Fusional 
Integrator for Radiation-hydrodynamic Systems in Tsukuba University).
The {\it FIRST} simulator is a large-scale hybrid PC cluster, where each node possesses a newly-developed 
board for gravity calculations, Blade-GRAPE. The Blade-GRAPE is 
composed of four GRAPE-6A chips\cite{FMK05}. The 
theoretical peak performance of one Blade-GRAPE board is 136.8 Gflops. Each board has 16MB of memory 
and can treat 260,000 particles simultaneously. The Blade-GRAPE is directly connected via PCI-X bus. 
Each server PC is equipped with a multi-port Gigabit Ethernet NIC that is connected to a special 
interconnection network using commodity Ethernet switches. 

The first version of Blade-GRAPE works with 32 bit and 33 MHz (PCI). Then, it has been improved to a 
64 bit and 100 MHz version (PCI-X), Blade-GRAPE X64. Using Blade-GRAPEs, we have constructed 
a hybrid PC cluster system composed of 256 nodes, the {\it FIRST} simulator (Fig. \ref{FIRST}).  
The system possesses 224 Blade-GRAPE X64 boards and 16 Blade-GRAPE boards. 
The host PC cluster node is a 2U-size server PC (HP ProLiant DL380 G4) that has dual Xeon processors 
in a SMP configuration. As a result, the total performance of the {\it FIRST} simulator is 36.1 Tflops, 
where the host PC cluster is 3.1 Tflops and the Blade-GRAPEs are 33 Tflops. 
All nodes are connected uniformly to each other via a multi-port Gbit ethernet interconnect switch. 
The total memory of the {\it FIRST} simulator is 1.6 TB. 
Also, the Gfarm Grid file system (http://datafarm.apgrid.org/index.en.html) is installed. With Gfarm,
a total storage of 89.2 TB is available as a seamless file server. 

In this paper, we present hydrodynamic and radiation hydrodynamic
simulations on the formation of first objects and first galaxies with the {\it FIRST} simulator.

\section{Collapse of First Objects by Dark Matter Cusps}

\begin{table}[b]
\caption{Models and mass resolution}
\centering
\begin{tabular}{ccc} 
\hline \hline
Model & $(N_{\rm DM}, N_{\rm b})$ & $(m_{\rm DM}, m_{\rm b})$ \\ 
 & & $[M_\odot]$ \\
\hline
R64 & $(64^3, 64^3)$ & $(24.1, 5.08)$ \\ 
R128 & $(128^3, 128^3)$ & $(3.01, 0.64)$ \\ 
R256 & $(256^3, 256^3)$ & $(0.38, 0.079)$ \\ 
R512 & $(512^3, 512^3)$ & $(0.047, 0.0099)$ \\ 
\hline
\label{table:models}              
\end{tabular}
\end{table}

The efficiency of the formation of first objects is extremely significant in that 
they are responsible for the reionization history in the universe and 
the cosmic chemical enrichment. 
Yoshida et al. (2003) \cite{Yoshida03} have shown 
that the minimum mass is a weak function of redshifts, which is  $\sim 7\times 10^5M_\odot$ in halo mass and 
$\sim 10^5M_\odot$ in baryonic mass. 
The minimum mass is basically determined by the thermal instability via 
hydrogen molecules (H$_2$). 
H$_2$ molecules form in the non-equilibrium processes with the catalysis of free electrons as
\begin{equation}
\begin{array}{lll}
e^{-} \, + \, {\rm H} \, & \rightarrow & \, {\rm H}^{-} \, + \, h \nu \\
{\rm H}^{-} \, + \, {\rm H} \, & \rightarrow & \, {\rm H} _2 \, 
+ \, e^{-}.
\end{array}
\label{H2-reaction}
\end{equation}
If the temperature exceeds $10^3$K around the number density of 1 cm$^{-3}$, 
the formation rate of H$_2$ molecules is increased, resulting in the H$_2$ abundance
of $y_{\rm H_2} \sim 10^{-3}$. Then, thermal instability occurs and
the temperature decreases down to $\sim 300$K with increasing density. 
Then, the runaway collapse of the cloud proceeds through gravitational instability. 
The simulations by Yoshida et al. (2003) demonstrated that for 
the halo mass below $7\times 10^5M_\odot$, the virial temperature cannot be raised up to $\sim 10^3$K, 
and therefore the thermal instability by H$_2$ cooling does not occur. 
The mass resolution of their simulations is $\approx 30M_\odot$.
Recently, Umemura et al. \cite{Umemura08a,Umemura10} have found
by simulations with higer mass resolution that 
significantly smaller mass haloes allow the collapse of primodial clouds 
via the H$_2$ cooling instability. The major difference from previous simulations is 
the fact that the mass resolution is much higher and not changed throughout the 
simulations, in contrast to the adaptive change of resolutions in previous works. 
Umemura et al. argued that the runaway in a smaller dark halo
could be induced by the cusp of dark halo. In other words,
the resolved cusp potential can raise the central temperature of cloud up to $\sim 10^3$K 
and leads to the thermal instability via H$_2$.
In this paper, to verify this conjecture, we compare the numerical results with
different mass resolution, and analyze the dark matter cusp-induced collapse.

\subsection{Set-up of Simulations}

The formation of first objects from primordial gas is driven by dark matter 
fluctuations and cooling by H$_2$ molecules. In order to investigate the 
dependence on dark matter potential, we perform high-resolution cosmological 
hydrodynamic simulations in a standard $\Lambda$CDM cosmology with 
the parameters of $(\Omega_\Lambda, \Omega_{\rm m}, \Omega_{\rm b}, h)
=(0.72, 0.24, 0.042, 0.73)$. 
First, we perform pure $N$-body simulations for dark matter in a larger box 
with a commoving linear scale of 1Mpc and find the highest density domain of (60kpc)$^3$. 
For this domain, the evolution of density fluctuations is recalculated including baryons
from the recombination epoch ($z=10^3$). 
We use a Particle-Particle-Particle-Mesh (P$^3$M) scheme for gravity force calculations, 
and the baryonic component is treated with the Smoothed Particle Hydrodynamics (SPH) method. 
Radiative heating/cooling and chemical reactions for e$^-$, H, H$^+$, H$^-$, H$_2$, H$_2^+$
are included, where helium and deuterium reactions are dismissed. 
The direct part of self-gravity is calculated with Blade-GRAPE in the {\it FIRST} simulator. 
The simulations do not use any zoom-in technique, but the mass resolution is uniform 
over the whole computational domain and throughout the evolution. 
The system mass is $6.3 \times 10^6 M_\odot$ 
in dark matter and $1.6 \times 10^6 M_\odot$ in baryonic matter. 
The models with different resolutions are summarized in Table \ref{table:models}. 

\subsection{Growth of dark matter cusps}

\begin{figure}[t]
\begin{center}
\includegraphics[width=14.5cm]{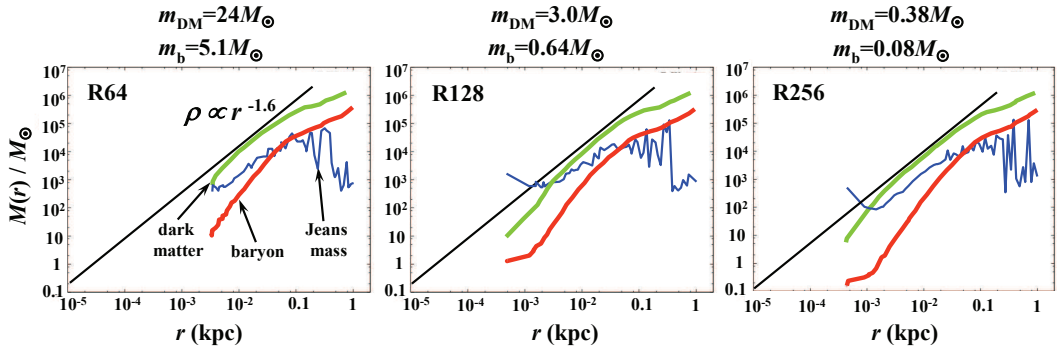}
  \caption{Cumulative mass of dark matter and baryons is shown 
against radius at $z=16$ for different mass resolution simulations. 
The left panel shows model R64, the middle panel does model R128, 
and the right panel does model R256. A straight line shows the mass 
corresponding to the density distribution of $\rho \propto r^{-1.6}$.
The Jeans mass including dark matter that is defined by (\ref{Jeans}) is also shown. 
}
\end{center}
\label{fig:cusp}
\end{figure}

\begin{figure}[t]
\includegraphics[width=14cm]{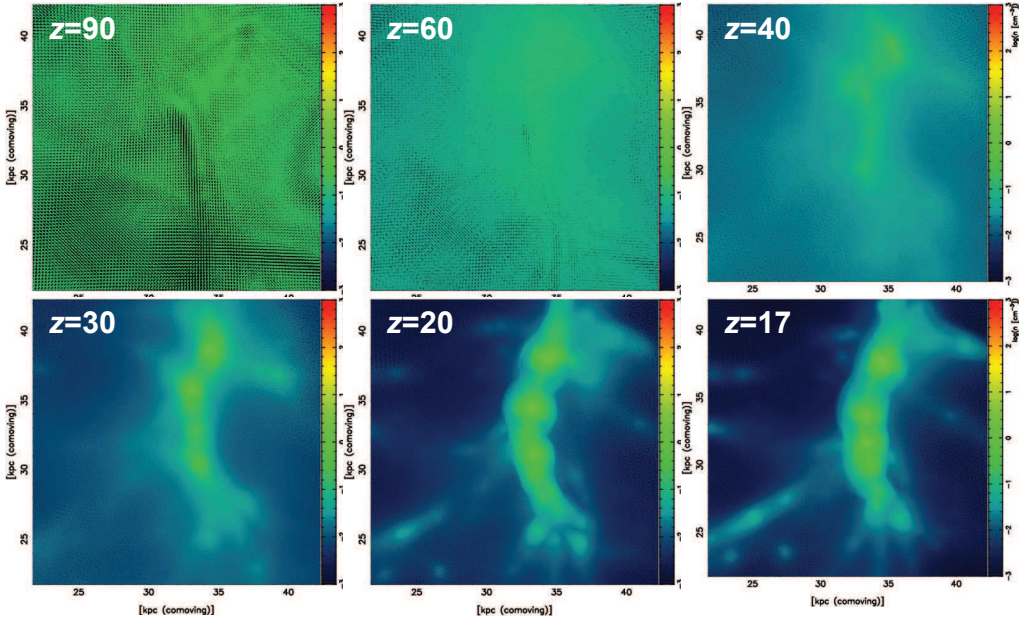}
\caption{Evolution of baryonic density fluctuations from $z=90$ to $z=17$
for model R256. Colors show the density according to the legend of colored bar.}
\label{fig:evolution}
\end{figure}

A dark matter fluctuation in the computational domain enters 
a non-linear stage around $z=30$. 
In the course of the non-linear evolution, dark matter forms a virialized halo, 
and simultaneously a central cusp develops. The growth of the cusp depends on the mass resolution.  
In Fig. \ref{fig:cusp}, the mass distributions are shown for different mass resolution
simulations. In model R256, the cusp intrudes to radii of $\sim 1$pc. 
The density distribution is well fitted by $\rho \propto r^{-1.6}$. 
In a low or intermediate resolution case (model R64 or R128), 
the cusp is smoothed out at innermost regions. 
The dark matter density profile with a cusp is often expressed by 
\begin{equation}
\rho(r)=\rho_s(r/r_s)^{-a}[1+(r/r_s)^b]^{(3-a)/b},
\end{equation} 
wehre $r_s$ is a characteristic inner radius and $\rho_s$ is the corresponding inner density. 
In the NFW profile \cite{NFW}, $(a, b)=(1, 1)$ and the mass enclosed within $r$ is 
$M(r)=4\pi r_s^3{{\rm ln}[1+(r/r_s)]-(r/r_s)/[1+(r/r_s)]}$ 
or in the Moore's profile \cite{Moore} $(a, b)=(1.5, 1.5)$ and the mass is 
\begin{equation}
M(r)=(8\pi /3)\rho_s r_s^3 {\rm ln}[1+(r/r_s)^{1.5}].
\label{eq:Mr}
\end{equation} 
The pure $N$-body simulations of dark matter hitherto performed have shown that
the cusp profiles are between the NFW profile and the Moore's one.
\cite{NFW,FM97,Moore,FKM04,Diemand08}
The profiles of cusps ($\rho \propto r^{-1.6}$) in the present simulations are 
close to the Moore's profile, although the preset results are slightly
steeper. (This may be the effect of dissipation of baryonic component.)
For a virialized halo with $\approx 10^4M_\odot$, 
the cusp mass is $\approx 10^3M_\odot$ 
if the concentration parameter is $c\equiv r_{\rm vir}/r_s \approx 10$ 
with the virial radius $r_{\rm vir}$. 
If the mass of dark matter particle is higher than $M_\odot$,  
a cusp with $\approx 10^3M_\odot$ is made of less than several hundred particles. 
The two-body relaxation proceeds in a timescale of $0.1N t_{\rm dyn}/{\rm ln} N$,  
where $N$ is the number of particles and $t_{\rm dyn}$ is the dynamical time.
If $N<10^3$, the two-body relaxation timescale is less than $8 t_{\rm dyn}$.
Thus, the dark matter cusp is inevitably erased by the two-body relaxation. 
Actually, in model R64 ($m_{DM}=24.1M_\odot$) or R128 ($m_{DM}=3.01M_\odot$),
the mass distribution decreases steeply toward the center. This meams
that the central cusps are smoothed out by the two-body relaxation.
On the other hand, in model R256 ($m_{DM}=0.38M_\odot$), a cusp grows
down to $\sim$1pc. Consequently, the central dark matter potential
becomes deeper.

In Fig. \ref{fig:cusp}, we also show the Jeans mass defined by 
\begin{equation}
M_J (r) \equiv {\rho _b}{\left( {\frac{{c_s^2}}{{G{{\bar \rho }_{tot}}}}} \right)^{3/2}} 
= {\rho _b}{\left( {\frac{{4\pi {r^3}c_s^2}}{{3G [ {{M_{\rm DM}}(r) + {M_{\rm b}}(r)} ] }}} \right)^{3/2}},
\label{Jeans}
\end{equation}
where $c_s$ is the local sound velocity, and $M_{\rm DM}$ and $M_{\rm b}$
are the cumulative mass of dark matter and baryons, respectively.
As seen clearly, for higher mass resolution the Jeans mass is reduced 
in innermost regions owing to the dark matter cusp. 
The central dark matter potential is responsible for
the temperature, which is directly related to the thermal instability
by H$_2$ molecules.

\subsection{Thermal instability induced by a dark matter cusp}

\begin{figure}[t]
\includegraphics[width=14cm]{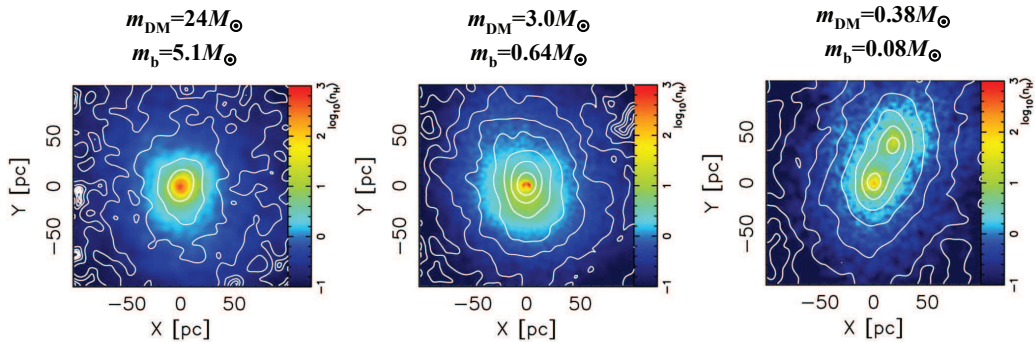}
\caption{Spatial density distributions of baryons and dark matter at $z=12$
are shown for different resolution simulations. 
The left panel shows model R64, the middle panel does model R128, 
and the right panel does model R256. 
Colors shows baryonic density following the attached legend, and 
contours show the levels of dark matter density.
}
\label{fig:contour}
\end{figure}

\begin{figure}[t]
\begin{center}
\includegraphics[width=10cm]{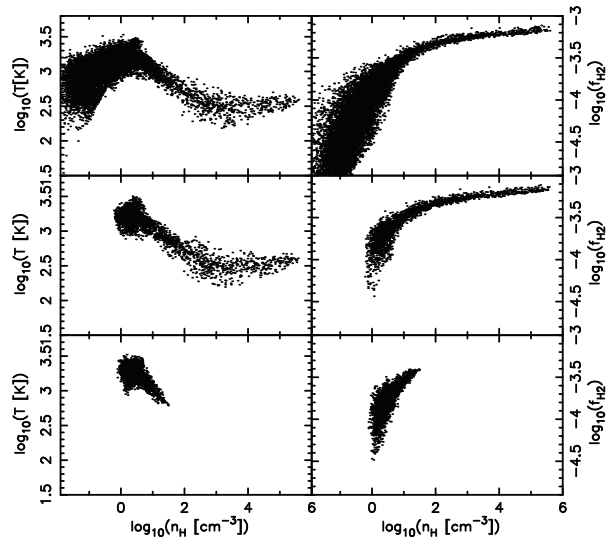}
\caption{Temperature ({\it left}) and H$_2$ fractions ({\it right}) 
versus barynon number density in model R256 are shown 
for all particles ({\it top panels}), the particles in the highest peak ({\it middle panels}), and
the particles in the second highest peak ({\it bottom panels}).
}
\label{fig:rho-T}
\end{center}
\end{figure}

In Fig. \ref{fig:evolution}, the evolution of baryonic density fluctuations 
induced by dark matter is shown from $z=90$ to $z=17$
for model R256. 
If the dark matter particle mass is less than  $1M_\odot$, the cusp with 
the density profile of  $r^{-1.6}$ is resolved down to $<10^3M_\odot$. 
In contrast to the previous works, multiple peaks can collapse induced 
by dark matter cusps. 
In Fig. \ref{fig:contour}, the density distributions at $z=12$ are compared among 
different resolutions. 
If the dark matter resolution is lower than 
$m_{\rm DM}=10M_\odot$, the cusp cannot be resolved sufficiently in a halo of $10^4 M_\odot$, 
and small-scale peaks readily merge into a larger peak, and consequently only one 
baryon condensation collapses within a halo.
This evolution is basically equivalent to the previous works. 
But, in model R256 ($m_{\rm DM}=0.38M_\odot$), a double peak forms, where the mass of each peak is 
$\approx 10^4M_\odot$ in dark matter and $\approx 10^3M_\odot$ in baryons.
The separation of peaks is roughly 60pc. 
Unless they undergo the thermal instability, the collapse of each peak bounces 
and the peaks may eventually merge into one larger peak.
Using eq (\ref{eq:Mr}), the baryonic gas temperature in the cusp region
can be evaluated as
\begin{equation}
T_{\rm cusp}=\frac{GM(r)}{r} \cdot \frac{\mu m_p}{k_B}
\simeq
1040{\rm{K}}\left( {\frac{{{\rho _s}}}{{1.6 \times {{10}^{ - 22}}
{\rm{g c}}{{\rm{m}}^{{\rm{ - 3}}}}}}} \right){\left( {\frac{{{r_s}}}{{10{\rm{pc}}}}} \right)^3}
{\left( {\frac{{{r_{}}}}{{10{\rm{pc}}}}} \right)^{-1}},
\label{eq:Tcusp}
\end{equation} 
where $\mu$ is the mean molecular weight, $m_p$ is the proton mass, and
$k_B$ is the Boltzmann constant.
Thus, if the cusp potential is resolved, the temperature can be raised up to
$\gtrsim 10^3$K, and then thermal instability through $H_2$ cooling can occur.
In Fig. \ref{fig:rho-T}, 
the temperature and H$_2$ fractions in model R256 are shown versus 
baryon number density. 
Top panels present those for all particles, 
middle panels for the particles in the highest peak, and
bottom panels for the particles in the 2nd highest peak.
This figure shows that 
H$_2$ fractions reach a level of several $10^{-4}$
around the density of $n_{\rm H}\sim 1 {\rm cm}^{-3}$ in the first and 2nd peaks.
Then, the temperature decreases via thermal instability in both peaks.
What is important is that 
although the halo mass of peaks is of the order of $\sim 10^4M_\odot$, 
the temperature reaches over $\sim 10^3$K around 
$n_{\rm H}\approx 1 {\rm cm}^{-3}$ owing to the deep potential of
dark matter cusps. 
Eventually,  dark haloes of $\sim 10^4M_\odot$ can produce primordial objects.
It means that the mass of first generation objects can be smaller 
by roughly two order of magnitude than in previous prediction. 
Hence, we conclude that the number of Pop III stars formed in an early universe 
can be significantly larger than in the previous prediction.

\begin{figure}[t]
\begin{center}
\includegraphics[width=12cm]{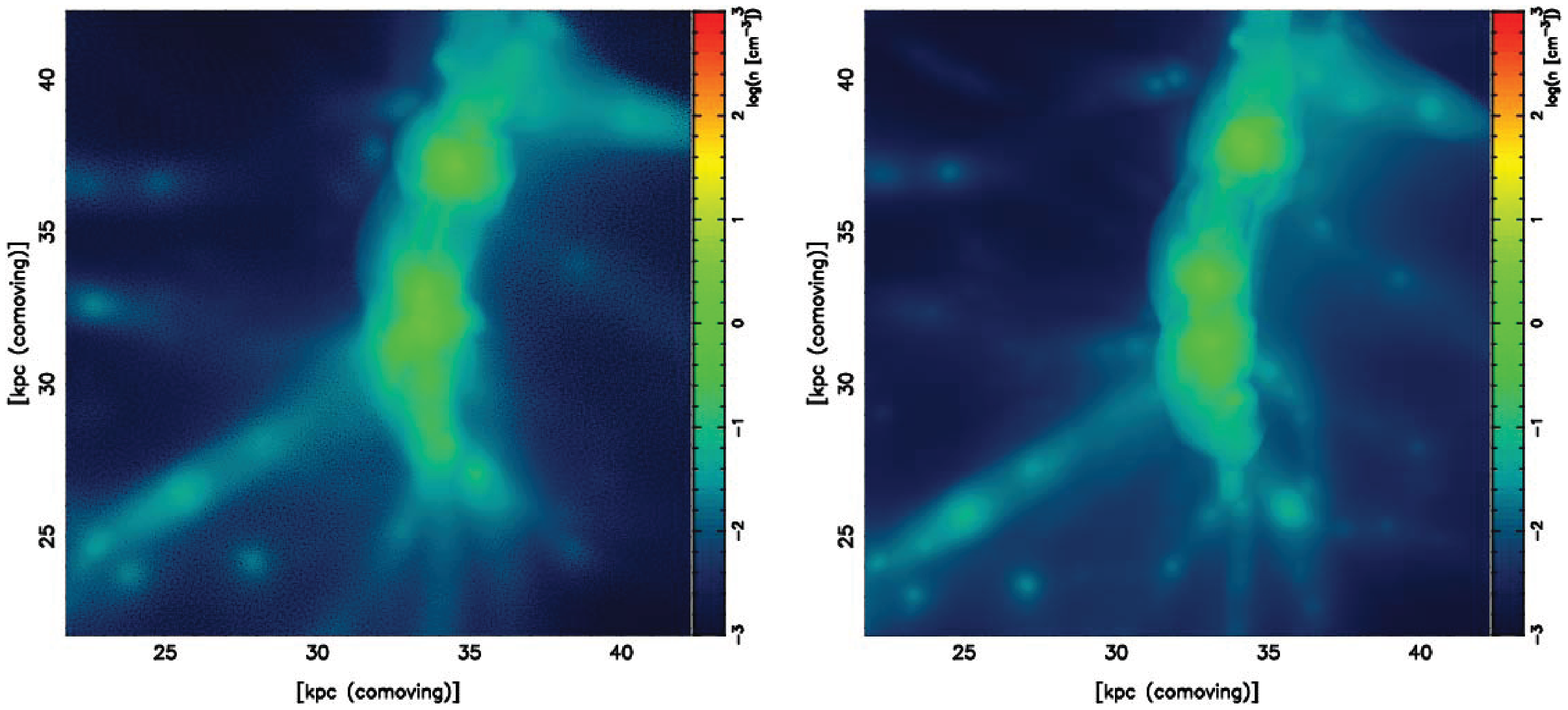}
\caption{Baryonic density distributions are compared at $z=15.6$ between
model R256 ({\it left}) and R512 ({\it right}). 
Colors shows baryonic density following the attached legend.
}
\label{fig:comparison}
\end{center}
\end{figure}

We have demonstrated that the resolution of dark matter cusps is 
crucial for the thermal instability. Here, we make the convergence 
test regarding the resolution to verify the present results. 
We perform a ultra-high resolution simulation (model R512).
In Fig. \ref{fig:comparison}, we compare the results between models R256 and R512.
We confirm that both give basically the same results. 
This means that the dark matter cusp resolved down to 1pc 
is important for thermal instability. The further inner cusp is not 
responsible for a significant increase of temperature. 

\section{Radiation Hydrodynamic Feedback in First Objects}

In the previous section, we have shown that two peaks with separation of
$\approx 60$pc collapse independently.
Although each peak could form a fairly massive star of $\gtrsim 40M_\odot$,
\cite{Hosokawa11}
there is a difference of free-fall time of $\sim 5\times 10^6$yr.
Hence, the first peak forms a massive Pop III appreciably
ealier than the 2nd peak, and then the first star irradiates
the second collapsing peak with strong UV radiation. 
Thus, whether the 2nd peak successfully forms a star depends 
upon the strength of radiative feedback from the first star.
The physical mechanism of radiation hydrodynamic feedback has been studied by
Susa \& Umemura (2006)\cite{SU06}, 
Susa et al. (2009)\cite{Susa09}, and 
Hasegawa et al. (2009) \cite{Hasegawa09}.
We show below the basic physics of radiation hydrodynamic feedback
and explore the survival of the 2nd peak under the feedback. 

\subsection{Basic Physics}

\begin{table}[b]
  \caption{Properties of Pop III stars}
  \begin{center}
  \begin{tabular}{llll}
  \hline
  Mass  & $T_{\rm eff}$ [K] $^{a)}$ & $\dot N_{\rm ion}$ [${\rm s^{-1}}$] $^{b)}$ & $L_{\rm LW}$[{\rm erg/s}] $^{c)}$ \\
  \hline
  $120M_{\odot}$ & $9.57\times 10^4$ & $1.069\times 10^{50}$ & $5.34 \times 10^{23}$ \\
  $80M_{\odot}$ & $9.33\times 10^4$ & $5.938\times 10^{49}$ & $3.05 \times 10^{23}$ \\
  $40M_{\odot}$ & $7.94\times 10^4$ & $1.873\times 10^{49}$ & $1.17 \times 10^{23}$ \\
  $25M_{\odot}$ & $7.08\times 10^4$ & $5.446\times 10^{48}$ & $3.94 \times 10^{22}$ \\
   \hline
\end{tabular}
\end{center}
{\scriptsize
~\\
a) Effective temperature \\
b) Number of ionizing photons emitted per second \\ 
c) Luminosity of photo-dissociation radiation at Lyman-Werner (LW) band 
}
\label{table:PopIII-stars} 
\end{table}

The properties of Pop III stars in the range of $25M_{\odot} \leq M_* \leq 120M_{\odot}$
are summarized in Table \ref{table:PopIII-stars}, which are taken from Schaerer (2002) \cite{Schaerer02}. 
Since Pop III stars emit strong ionizing and photo-dissociating radiation,
the radiation hydrodynamic feedback is regulated by the propagation
of an ionizing front and the shielding from photo-dissociating radiation.
If a collapsing core is irradiated by an ionizing source located
at a distance $D$, the propagation speed of the ionization-front (I-front) 
in the core is given by
\begin{equation}
v_{\rm IF}=
21~ {\rm km~s^{-1}}
\left(\frac{\dot{N}_{\rm ion}}{10^{50}{\rm s^{-1}}}\right)
\left(\frac{D}{20{\rm pc}}\right)^{-2}
\left(\frac{n_{\rm core}}{10^3{\rm cm^{-3}}}\right)^{-1},
\end{equation}
if recombination is neglected, 
where $\dot{N}_{\rm ion}$ is the number of ionizing photons per
unit time and $n_{\rm core}$ is the number density in the cloud core. 
The sound speed in regions cooled by H$_2$ 
is $a_1 \approx 1{\rm km~s^{-1}}$, 
while that in the ionized regions is $a_2 \approx 10{\rm km~s^{-1}}$. 
If the density of the cloud core is low or the ionizing radiation flux is strong, then 
$v_{\rm IF} > 2a_2$ and therefore the I-front becomes R-type. 
If we focus on a core collapsing in a self-similar fashion \cite{ON98},
the core size $r_{\rm core}$ is on the order of 
$a_1 t_{\rm ff}$, where $t_{\rm ff}$ 
($\simeq \sqrt{\pi/G\rho_{\rm core}}$) is the free-fall time. 
Then, the propagation time of R-type front across the core satisfies
$
t_{\rm IF}\equiv r_{\rm core}/v_{\rm IF} <
 (a_1/2a_2) t_{\rm ff} < t_{\rm ff}.
$
This means that an R-type front sweeps the core before the core 
collapses in the free-fall time. Thus, the core is likely to undergo
photo-evaporation. 
On the other hand, if the density of the cloud core
is high enough or the source distance is large, then
$v_{\rm IF} < a_1^2/2a_2$ and a D-type I-front emerges. 
The propagation time of a D-type front across the core satisfies
$
t_{\rm IF}  > (2a_2/a_1) t_{\rm ff} > t_{\rm ff}.
$
Thus, the core can collapse before the I-front sweeps the core. 

However, the above arguments are based on the assumption that
the ionizing photon flux does not change during the propagation
of I-front. The core could be self-shielded from the ionizing
radiation if 
$\dot{N}(\pi r_{\rm core}^2/4\pi D^2)<4\pi r_{\rm core}^3 
n_{\rm core}^2 \alpha_{\rm B}/3$, where $\alpha_{\rm B}$
is the recombination coefficient to all excited levels of hydrogen.
The critical density for self-shielding is given by
\begin{eqnarray}
&\;& n_{\rm shield} \simeq \left(\frac{3\dot{N}_{\rm ion}}
{16\pi D^2 a_1 \alpha_{\rm B}} \sqrt{\frac{G m_{\rm p}
}{\pi}}\right)^{2/3}\nonumber\\
&=& 5.1~ {\rm cm^{-3}}
\left(\frac{\dot{N}_{\rm ion}}
{10^{50}{\rm s^{-1}}}\right)^{2/3}\left(\frac{D}
{20{\rm pc}}\right)^{-4/3}
\left(\frac{a_1}{1{\rm km~ s^{-1}}}\right)^{-2/3}. \label{eq:shield}
\end{eqnarray}
If $n_{\rm core}> n_{\rm shield}$, the ionizing photon flux 
diminishes significantly during the I-front propagation.
Hence, even if the I-front is R-type on the surface of
the cloud core, the front can change to M-type accompanied by a shock,
and eventually to D-type inside the core \cite{Kahn54}. 

In contrast to ionizing radiation, H$_2$ dissociating radiation in 
LW band (11.26-13.6 eV) is less shielded \cite{DB96}. 
The self-shielding of LW band flux ($F_{\rm LW}$) is expressed by
\begin{equation}
F_{\rm LW} = F_{\rm LW,0} f_{\rm sh}
\left( N_{\rm H_2,14 } \right) \label{LW}
\end{equation}
where $ F_{\rm LW,0}$ is the incident flux, 
$ N_{\rm H_2,14}= N_{\rm H_2}/10^{14} {\rm cm^{-2}}$
is the normalized H$_2$ column density, and
\begin{equation}
f_{\rm s}(x) = \left\{
\begin{array}{cc}
1,~~~~~~~~~~~~~~x \le 1 &\\
x^{-3/4},~~~~~~~~~x > 1 &
\end{array}
\right.
\label{shield-function}
\end{equation}
Hence, if the column density of H$_2$ molecules ahead of I-front is high enough,
H$_2$ dissociating radiation can be shielded.
Since H$_2$ molecules form with the catalysis of free electrons,
the shielding of dissociating radiation is intimately coupled with the propagation of I-front.

\subsection{START: Accelerated ratiation hydrodynamic scheme}

\begin{figure}[t]
\begin{center}
\includegraphics[width=12cm]{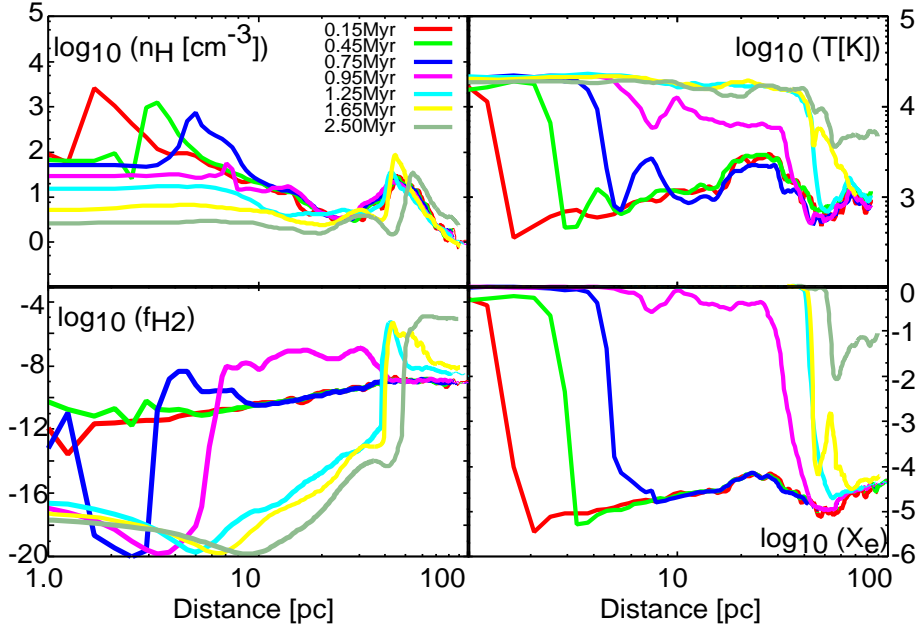}
\caption{Time variations of gas density, temperature, H$_2$ fractions, and 
ionization degree along the line connecting two peaks
are shown at the epochs from 0.15Myr to 2.5Myr. The distance is measured
from the first peak. The 2nd peak is located at $\sim 60$pc
}
\label{fig:feedback-evol}
\end{center}
\end{figure}

In order to perform the radiation hydrodynamic (RHD) simulations more effectively,
recently we have developed a novel RHD code called {\it START} (SPH with Tree based 
Accelerated Radiation Transfer), which is designed to solve the transfer of UV photons 
from numerous sources \cite{rf:HU10}. 
In {\it START}, the optical depth between a radiation source and 
an SPH particle is evaluated with the same method as in the {\it RSPH} code\cite{Susa06}. 
In contrast to the corresponding method proposed by Kessel-Deynet \& Burkert (2000)\cite{rf: KB00},
the evaluation of the optical depth between a radiation source and a target SPH particle is 
performed only once by using information of its upstream particle. 
Thus, the cost for each ray-tracing turns out to be proportional to $N_{\rm SPH}$, 
where $N_{\rm SPH}$ is the numbers of SPH particles.
In addition to the reduction of the cost for each ray-tracing, 
the effective number of radiation sources for each target SPH particle is diminished 
by utilizing an oct-tree structure of radiation sources. 
Similar to the Barnes-Hut tree method \cite{rf:BH86} that is frequently used 
for calculating the gravitational force, if a cell containing some radiation sources 
is far enough from a target SPH particle, all radiation sources in the cell is 
regarded as a virtual luminous source. 
As a result, the computational cost to evaluate optical depths from all radiation sources 
to all SPH particles is proportional to $O(N_{\rm SPH}\log N_{s})$. 
With this code, we can precisely solve the transfer of recombination photons, 
which is often treated with the on-the-spot approximation. 
Another important and useful feature of {\it START} is that the spatial resolution is adaptively 
enhanced in high density regions, since SPH particles are directly used for the radiative 
transfer grids. 
Here, using  {\it START}, we consistently solve the gas and 
dark matter dynamics coupled with the radiative 
transfer of UV photons and non-equilibrium chemical reactions for 6 species: 
${\rm e^{-}, H^+, H, H^-, H_2, \, {\rm and} ~H_2^+}$. 

\subsection{Radiation hydrodynamic feedback}

\begin{figure}[t]
\begin{center}
\includegraphics[width=10cm]{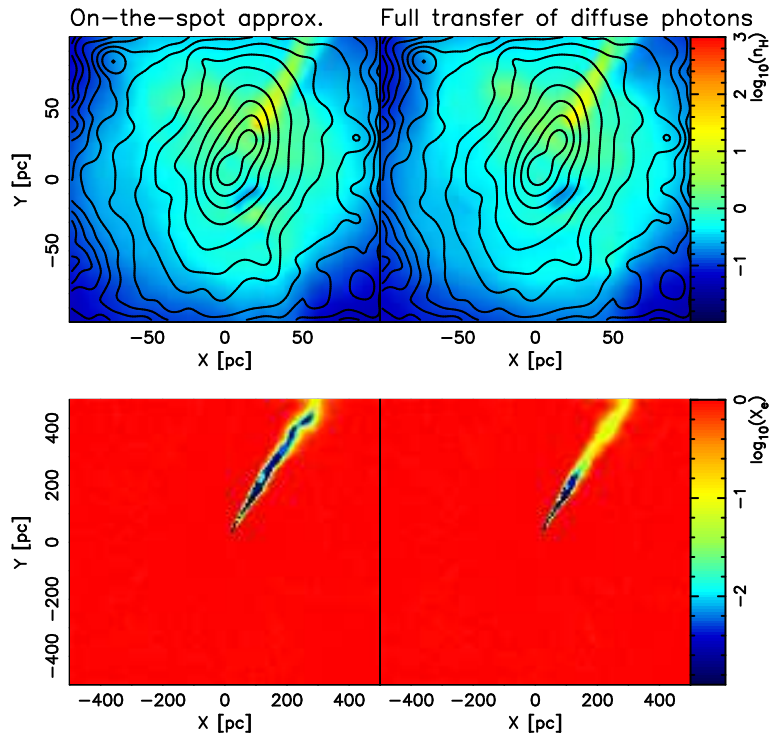}
\caption{Snapshots for the gas density ({\it upper row}), and 
the ionization degree ({\it lower row}) at 2.5Myr. 
The initial distribution at $t=0.0$Myr corresponds to the rightmost panel of Fig. \ref{fig:contour}.
The origin, (x,y)=(0,0) [pc], is set to be the position where the highest peak was originally located,
and the 2nd peak was originally located at (x,y)=(20,50) [pc].
The contours show the dark matter distributions. The left panels are the
results under the on-the-spot approximation, while the right panels show the 
results of full radiation-hydrodynamic simulations solving the transfer of
diffuse photons.
}
\label{fig:feedback-comp}
\end{center}
\end{figure}

\begin{figure}[t]
\includegraphics[width=14cm]{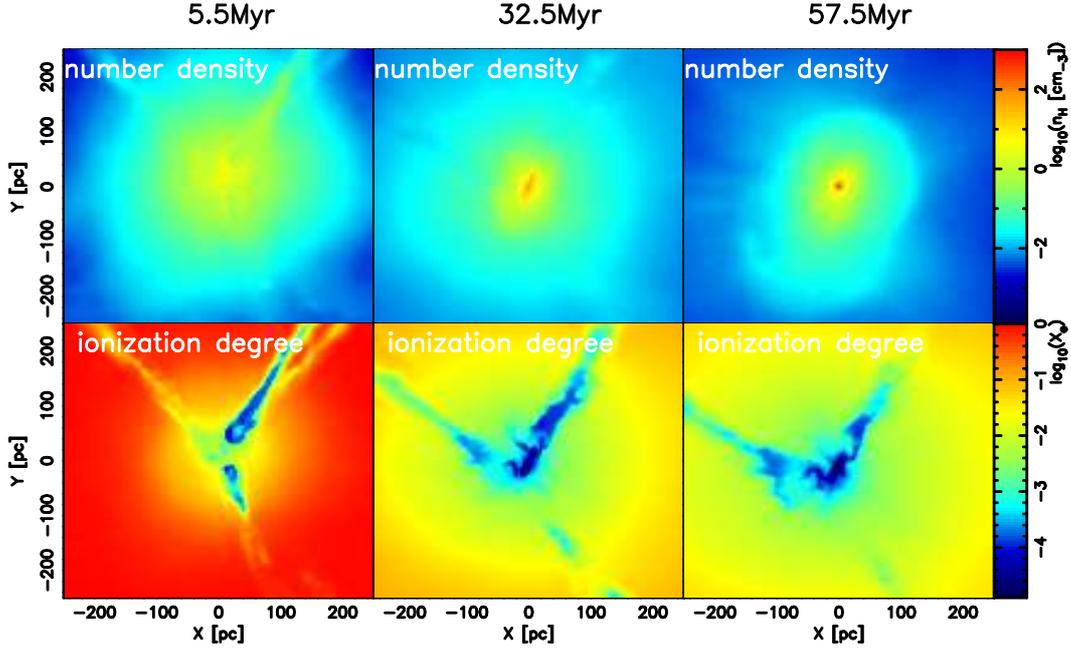}
\caption{The late-phase evolution of the 2nd peak is shown 
at 5.5Myr, 32.5Myr, and 57.5Myr. 
The origin, (x,y)=(0,0) [pc], is set to be the position where the highest peak was originally located,
and the 2nd peak was originally located at (x,y)=(20,50) [pc].
The gas density is shown in the upper row, and the ionization degree 
is shown in the lower row. 
}
\label{fig:feedback}
\end{figure}

We use the results obtained in the previous section, where
two dark matter haloes host gas clumps, and the separation between the halos is $\sim$60pc. 
The whole simulation box size is 2kpc (physical) on a side. The particle mass is $0.08M_{\odot}$ 
for an SPH particle, and $0.38M_{\odot}$ for a DM particle, respectively. We assume 
a source Pop III star with a mass of $120M_{\odot}$, and place it at the highest density peak.  
Then, we start the RHD simulations, where not only direct UV photons from the source 
Pop III star but also diffuse photons produced via recombination processes are considered. 
After the lifetime of the source star ($t=2.5{\rm Myr}$), direct UV photons from the source star is 
turned off, but the transfer of diffuse photons is continuously solved. 
After the source star dies, the supernova explosion may make an impact on the
collapsing core, which is not treated in this paper and will be explored elsewhere.

The evolution from 0.15Myr to 2.5Myr obtained by the radiation hydrodynamic simulation
is presented in Fig. \ref{fig:feedback-evol}, where time variations of gas density, temperature, H$_2$ fractions, 
and ionization degree along the line connecting two peaks are shown. The distance is measured
from the first peak, and the 2nd peak is originally located at $\sim60$pc.
In the early stage before $t=0.75$Myr, a shock precedes a D-type I-front. 
Before the irradiation of UV, the H$_2$ fraction at the 2nd peak is 
several $10^{-4}$, as shown in Fig. \ref{fig:rho-T}. 
However, as shown in Fig. \ref{fig:feedback-evol}, 
the the H$_2$ fraction at the 2nd peak is reduced to a level of $10^{-9}$ due to 
photo-dissociating radiation, although the H$_2$ fraction near the I-front is slightly enhanced.
After this stage, the I-front changes into R-type. 
The evolution until this early stage is very similar to the corresponding results 
shown by Kitayama et al. (2004) \cite{rf:Kitayama04} or Yoshida et al. (2007) \cite{rf:Yoshida07}. 
However, unlike these previous studies, the I-front changes into D-type again 
at $t\sim1.25$Myr due to the presence of the 2nd peak. 
Then, an H$_2$ shell with an H$_2$ fraction of $\sim 10^{-4}$ forms ahead 
of the 2nd peak thanks to the catalysis of free electrons. 
This H$_2$ shell shields photo-dissociating radiation significantly, and consequently 
the H$_2$ fraction at the 2nd peak is restored to
$\sim 10^{-4}$. Simultaneously, the envelope of the 2nd peak is stripped by 
the shock associated with the I-front. 

The ionization around the 2nd peak is dependent on the treatment of UV radiation transfer.
In Fig. \ref{fig:feedback-comp}, the results under the on-the-spot approximation are 
compared to those of full RHD simulations solving the transfer of diffuse photons at the epoch of 2.5Myr. 
As shown in the top panels of Fig. \ref{fig:feedback-comp}, the diffuse UV radiation 
hardly affects the gas distribution in the 2nd peak,
because the mean free path of ionizing photons is quite short there. 
Thus, the impact of the diffuse UV radiation on the 2nd peak is dimunitive. 
On the other hand, the difference of the ionized fractions in the outer envelope is 
obvious behind the 2nd peak.
The full RHD simulations show that diffuse UV radiation can ionize
the gas behind the 2nd peak, since the photon mean free path is relatively long in this region.
Therefore, solving the transfer of diffuse UV radiation is of great significance 
in computing the ionization of the low-density gas, e.g., intergalactic medium.

\begin{figure}[t]
\begin{center}
\includegraphics[width=5cm]{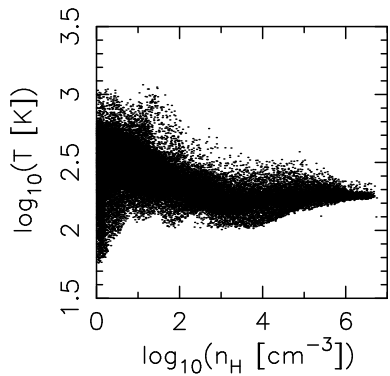}
\caption{Temprature versus density at the end of the simulation. 
}
\label{fig:rhoT-feedback}
\end{center}
\end{figure}

The later evolution of the 2nd peak is determined by the interplay of
the shielding from photo-dissociating radiation and the stripping by an ionization-front shock.
In Fig. \ref{fig:feedback}, the later evolution in full RHD simulations is
presented at 5.5Myr, 32.5Myr, and 57.5Myr, where the distributions of density and 
ionization degree are shown.
After $t=$2.5Myr, no UV radiation is emitted from the source.
Thus, the recombination gradually increases the neutral fractions. 
But, diffuse UV radiation from the recombination ionizes the shadowed region 
behind the 2nd peak. Although the central density of the 2nd peak decreases 
via the photoionization and a shock associated with the ionization-front,
the 2nd peak collapses eventually.  
In Fig. \ref{fig:rhoT-feedback}, 
the the density of the 2nd peak is shown against temperature.
This figure shows that the 2nd peak continues to collapse after the thermal
instability, irrespective of the radiative feedback. 

\subsection{Analytic Criteria}

Here, we derive analytic criteria for the feedback, based on the argument 
by Hasegawa et al. (2009)\cite{Hasegawa09}. 
Susa (2007)\cite{Susa07} has explored the photo-dissociation feedback of a Pop III  star 
with $120M_{\odot}$ on a neighboring prestellar core by RHD simulations 
without ionizing radiation. Susa (2007) has found that a condition for the collapse of a neighboring core is 
approximately determined by $t_{\rm dis}=t_{\rm ff}$, where $t_{\rm dis}$ is the 
photo-dissociation timescale in the core and $t_{\rm ff}$ is the free-fall timescale. 
Using this condition, the critical distance $D_{\rm cr,d}$, 
above which a neighboring core is able to collapse, is given by 
\begin{eqnarray}
	D_{\rm cr,d}& =& 147 {\rm pc}
	\left(\frac{L_{\rm LW}}{5\times 10^{23}{\rm erg\;s^{-1}}}\right)^{\frac{1}{2}}
	\left(\frac{n_{\rm c}}{10^{3}{\rm cm^{-3}}}\right)^{-\frac{7}{16}} \nonumber \\
	&&\times \left(\frac{T_{\rm c}}{300{\rm K}}\right)^{-\frac{3}{4}},  \label{Dcrd}
\end{eqnarray}
where $L_{\rm LW}$, $n_{\rm c}$, and $T_{\rm c}$ are the 
LW luminosity of the source star, the number density of the core, and 
the temperature of the core, respectively. 
This equation involves the self-shielding effect by the core. 
The dependence on the core temperature basically 
originates in the core radius ($\propto T_{\rm c}^{1/2}$) 
and a $\rm H_2$ formation rate in the core ($\propto T_{\rm c}$). 
Hence, the self-shielding effect is weaker for a lower core temperature. 
The boundary between the collapse irrespective of ionizing radiation 
and that with the aid of ionizing radiation can be determined by $D_{\rm cr,d}$. 
In addition, $D_{\rm cr,d}$ gives a good estimate 
for less massive source star cases, because the ionizing radiation is 
relatively weak for less massive stars. 

If ionizing radiation is included,
we should also incorporate the shielding effect by an $\rm H_2$ shell,
which is formed with the catalysis of free electrons. 
Here, we derive a new criterion including this effect. 
Since a cloud collapses in a self-similar fashion before UV irradiation, 
the density profile of the outer envelope in the cloud is well expressed as 
\begin{equation}
	n(r) = n_{\rm c}\left(\frac{r_{\rm c}}{r}\right)^2, 
\end {equation}
where $r_{\rm c}$ is the core radius which roughly corresponds to 
the Jeans scale; 
\begin{equation}
	r_{\rm c} = \frac{1}{2}\sqrt{\frac{\pi k_B T_{\rm c}}{Gm_p^2n_{\rm c}}}, 
\end{equation} 
where $m_{\rm p}$ denotes the proton mass. 
Assuming that the thickness of the $\rm H_2$ shell is determined by 
the amount of ionized gas in the envelope 
and the $\rm H_{2}$ fraction in the shell is constant,   
the $\rm H_2$ column density of the shell $N_{\rm H_2,sh}$ is given by 
\begin{equation}
	N_{\rm H_2,sh}=\int_D^{D_{\rm sh}}y_{\rm H_2, sh} n(r) dr
	=y_{\rm H_2,sh}n_{\rm c}r_{\rm c} ^2 \frac{D-D_{\rm sh}}{DD_{\rm sh}}, \label{NH2sh1}
\end{equation}
where $D_{\rm sh}$ and $y_{\rm H_2,sh}$ 
are the distance between the cloud core and the ${\rm H_2}$ shell, 
and the $\rm H_2$ fraction in the shell, respectively.   
Here, $D_{\rm sh}$ is set to be the position where the number of recombination 
per unit time in the ionized region around a source star balances the number 
rate of incident ionizing photons, since the $\rm H_2$ shell appears ahead 
of the ionization front. 
Hence, $D_{\rm sh}$ satisfies  
\begin{eqnarray}
	\frac{\dot{N}_{\rm ion}\pi D_{\rm sh}^2}{4\pi (D-D_{\rm sh})^2 }
	&=&2\pi \alpha_{\rm B}\int^{D_{\rm sh}}_{D}n(r)^2r^2dr \nonumber \\
	&=&2\pi \alpha_{\rm B }n_c^2 r_c^4\frac{D-D_{\rm sh}}{DD_{\rm sh}}. \label{Nion}
\end{eqnarray}
Using equation (\ref{NH2sh1}) and (\ref{Nion}), we obtain 
\begin{equation}
	N_{\rm H_2,sh}=y_{\rm H_2,sh} n_{\rm c}^{\frac{1}{3}}r_{\rm c}^{\frac{2}{3}}
	D^{-\frac{2}{3}}\left(\frac{\dot{N}_{\rm ion}}{8\pi \alpha_{\rm B}}\right)^{\frac{1}{3}}.
	 \label{NH2sh2}
\end{equation}
Because of the intense LW radiation, the $\rm H_2$ abundance at the shell is in chemical 
equilibrium. Therefore, $y_{\rm H_2,sh}$ is given by
\begin{equation}
	y_{H_2,sh}=\frac{n(D_{\rm sh})y_{\rm e,sh} k_{\rm H^-}}{k_{\rm dis}}, 
\end{equation}
where $y_{\rm e,sh}$ is the electron fraction at the $\rm H_2$ shell 
and $k_{\rm H^-}$  is the reaction rate of the $\rm H^-$ process by (\ref{H2-reaction}). 
In this case, we should consider the self-shielding effect by the shell itself. 
As a result, these rates are 
\begin{equation}
	k_{\rm H^-} = 1.0\times 10^{-18} T_{\rm sh} {\rm cm^{-3}s^{-1}}, 
\end{equation}
\begin{equation}
	k_{\rm dis} = 1.13 \times 10^8 F_{\rm LW,sh} 
	f_{\rm s}\left(\frac{N_{\rm H_2,sh}/2}{10^{14}\rm cm^{-2}}\right) \rm s^{-1}, \label{kdis}
\end{equation}
where $T_{\rm sh}$ and $F_{\rm LW,sh}$ are the temperature at the shell, and 
the LW flux from the star in the absence of shielding effect,  
$F_{\rm LW,sh}=L_{\rm LW}/4\pi(D-D_{\rm sh})^2$. 
$f_{\rm s}$ is the self-shielding function, which is given by
(\ref{shield-function}).  
Combining equations (\ref{NH2sh2})-(\ref{kdis}) 
with assumption of $y_{\rm e,sh}=0.05$ and $T_{\rm sh}=2000$K 
as shown in the present numerical results, we have 
 \begin{eqnarray}
	y_{\rm H_2,sh}
	&=&1.0\times 10^{-6}\left(\frac{\dot{N}_{\rm ion}}{10^{50}{\rm s^{-1}}}\right)^{\frac{11}{3}}
	\left(\frac{L_{\rm LW}}{5\times10^{23}{\rm erg\;s^{-1}}}\right)^{-4}\nonumber \\
	&&\times \left(\frac{T_{\rm c}}{300{\rm K}}\right)^{-\frac{1}{3}}
	\left(\frac{D}{40{\rm pc}}\right)^{\frac{2}{3}},
\end{eqnarray}
\begin{eqnarray}
	N_{\rm H_2,sh}&=&5.8\times10^{14}\left(\frac{\dot{N}_{\rm ion}}{10^{50}{\rm s^{-1}}}\right)^4
	\nonumber \\
	&& \times \left(\frac{L_{\rm LW}}{5\times 10^{23}{\rm erg\;s^{-1}}}\right)^{-4}{\rm cm^{-2}}. 
	\label{NH2sh3}
\end{eqnarray}
It should be noted that $N_{\rm H_2,sh}$ is independent of 
the core temperature $T_{\rm c}$, but is determined solely by the ratio of $\dot{N}_{\rm ion}$ to 
$L_{\rm LW}$ with strong dependence. 

Multiplying $L_{\rm LW}$ in equation (\ref{Dcrd}) by 
\begin{equation}
f_{\rm s,sh}\equiv f_{\rm s}\left(\frac{N_{\rm H_2,sh}}{10^{14}{\rm cm^{-2}}}\right),
\end{equation} 
we obtain the critical distance as  
\begin{eqnarray}
	D_{\rm cr,sh}& =& 147 {\rm pc}
	\left(\frac{L_{\rm LW}f_{\rm s, sh}}
	{5\times 10^{23}{\rm erg\;s^{-1}}}\right)^{\frac{1}{2}}
	\left(\frac{n_{\rm c}}{10^{3}{\rm cm^{-3}}}\right)^{-\frac{7}{16}} \nonumber \\
	&&\times \left(\frac{T_{\rm c}}{300{\rm K}}\right)^{-\frac{3}{4}},  \label{Dcrsh}
\end{eqnarray}
in which both shielding effects by the core and the $\rm H_2$ shell are taken into account. 
In particular, if $N_{\rm H_2,sh}$  is larger than $10^{14} \rm cm^{-2}$, 
the critical distance can be expressed as 
\begin{eqnarray}
	D_{\rm cr,sh} &=&78.8 {\rm pc}
	\left(\frac{L_{\rm LW}}
	{5\times 10^{23}{\rm erg\;s^{-1}}}\right)^{2}\left(\frac{\dot{N}_{\rm ion}}{10^{50}{\rm s^{-1}}}
	\right)^{-\frac{3}{2}}\nonumber \\
	&&\times \left(\frac{n_{\rm c}}{10^{3}{\rm cm^{-3}}}\right)^{-\frac{7}{16}}
	\left(\frac{T_{\rm c}}{300{\rm K}}\right)^{-\frac{3}{4}}. \label{Dcrsh2}
\end{eqnarray}
Based upon equation (\ref{NH2sh3}), the shielding effect by the shell becomes 
weaker as $\dot{N}_{\rm ion}/L_{\rm LW}$ decreases. 
Equation (\ref{Dcrsh2}) gives a criterion of distance, above which 
a cloud core can collapse through the shielding of photo-dissociation radiation
by an $H_2$ shell. 
The present results of radiation hydrodynamic simulations show that
a cloud located at 60pc collapses under the UV feedback.
This is broadly consistent with this analytic criterion. 
In the present simulations, the peak separation was $\sim$60pc.
Whether this separation is a typical value has not been revealed
in the present analysis. It needs the case study using more realizations
of random fluctuations, which is left in the future work.

\section{Fragmentation of a First Protostellar Disk and Radiative Feedback}

As stated in the Introduction, theoretical efforts in
the last decade revealed that very first stars are more massive than stars
forming in present-day galaxies. A few years before, it was believed
that they are very massive ($> 100 M_\odot$). However, recent numerical
studies using the sink particle technique suggest that owing to
the fragmentation in an accretion phase, they could be
$O(10)\times M_\odot$ or less, and eventually they form a multiple stellar system
\cite{Stacy10,Clark11,Greif11}. In some cases, it might be also possible for
sub-solar mass first stars to form from such fragments.  
In addition, recent two dimensional radiation hydrodynamic
simulations revealed that UV radiative feedback from 
first protostars is quite important to quench the mass accretion onto
the first protostar \cite{Hosokawa11}.

In this framework, we incorporate the UV radiative
feedback effects into the three-dimensional chemo-hydrodynamical
simulations, in order to follow the fragmentation of an accreting gas
disk in realistic circumstances.

\begin{figure}[b]
\begin{center}
\includegraphics[angle=0,width=10cm]{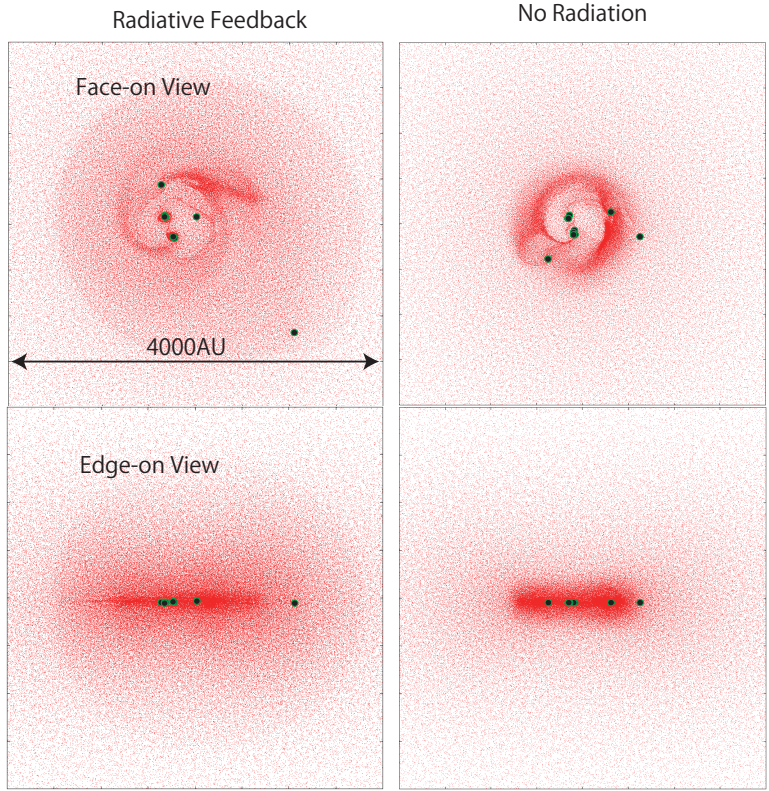}
\caption{Snapshots at 5500yrs after the first sink formation are
 shown. Red dots represent SPH particles, while black dots are sink
 particles. Left panels (face-on/edge-on view) are the results with
 radiative feedback, while the right panels without radiative feedback.} \label{fig:dots}
\end{center}
\end{figure}

\subsection{Numerical method with sink particles}

\begin{figure}[t]
\begin{center}
\includegraphics[angle=0,width=14.5cm]{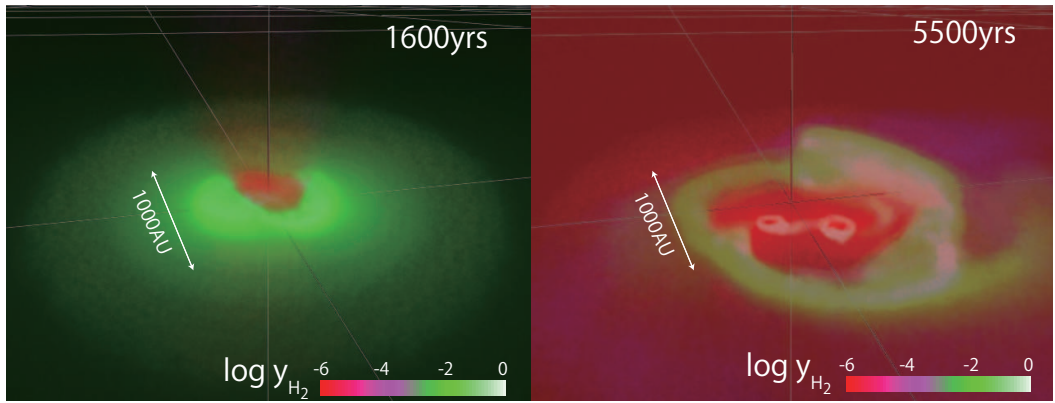}
\caption{Bird-eye view of the gas disk around the first sink at two
 epochs. Transparency describes the gas density, while the color shows
 the logarithmic fraction of H$_2$ molecules ($y_{\rm H_2}$). At earlier epoch
 (1600yrs), conical zone at the center is highly dissociated (red gas),
 while the other region is still unaffected by the central source. On
 the other hand, at a later epoch (5500yrs), only dense cold streams
 (white/green gas) orbiting around the center sustain significant amount
 of H$_2$, while the other less dense region is totally dissociated.} \label{fig:H2map}
\end{center}
\end{figure}

We perform numerical experiments using the {\it RSPH} code
developed by ourselves \cite{Susa06}. 
We set the the initial condition of a cloud to be 
a Bonner-Evert sphere with the density of $10^{4}$cm$^{-3}$ and the temperature of
$200$K, at the ``loitering'' phase of a collapsing primordial prestellar cloud. 
We add rigid rotation with angular velocity of $\rm{\Omega}=2\times 10^{-14} {\rm s}^{-1}$, which is
comparable to the rotation in cosmological simulations\cite{Yoshida03}.

In order to follow the accretion phase, we introduce the sink particle
technique in the {\it RSPH} code. In the case where the density 
at the position of an SPH particle exceeds $10^{13}{\rm cm^{-3}}$, 
it is regarded as a sink particle. Also, particles that enter
a sphere of radius 20AU centered on each sink particle 
are absorbed in the sink particle if they are gravitationally bounded.
In the present simulations, the mass of the SPH particle is $6\times
10^{-4}M_\odot$. Following previous studies\cite{bate_burket97}, the corresponding mass resolution
is $\sim 0.1 M_\odot$, which is well below the Jeans mass of the disk
at $1000$K, and that allows us to investigate the gravitational fragmentation
of the disk.
In the present work, we focus on the effects of Lyman-Werner radiation from
the first protostar on the fragmentation of the disk, and the resultant
accretion rate onto the first protostars. 
We solve non-equilibrium processes for H$_2$ formation. 
The transfer of the Lyman-Werner radiation is solved utilizing the self-shielding function \cite{rf:WH}.
The luminosity and the effective temperature of the protostars are
assessed by regarding the masses and the mass accretion rates of the sink
particles as those of the protostars. In paticular, we interpolate the
tables of luminosity and temperature generated by Hosokawa et al.\cite{rf:HYO}.
Although the effects of photoionization are also important,
we could not treat them due to the lack of spacial resolution.
We leave them in future works.

\begin{figure}[t]
\begin{center}
\includegraphics[angle=0,width=7cm]{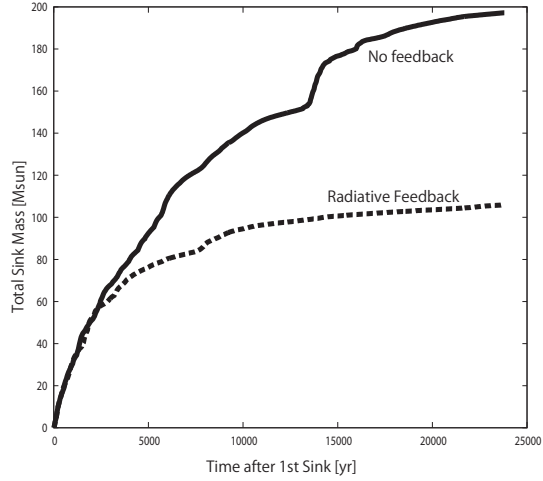}
\caption{Time evolution of total sink mass. Two curves correspond to
 the cases with/without radiative feedback.} \label{fig:total_sink_mass}
\end{center}
\end{figure}

\subsection{Radiative feedback in a first protostellar disk}

Fig. \ref{fig:dots} shows the numerical results, where the snapshots 
of distributions are presented at 5500yrs . Left panels
show the edge-on/face-on views of the gas disk in the inner (4000AU)$^3$ box of
the simulated region with radiative feedback, whereas the right panels
for the case without radiative feedback. In both cases, we observe that several
sink particles are formed (black circles), while the gas particle (red
dots) distributions are more extended in a case with radiative feedback.
This difference is a consequence of the H$_2$ photo-dissociation process.
Two panels in Fig. \ref{fig:H2map} show the bird-eye views of the gas disk around the
protostars at two epochs ($1600$yrs and $5500$yrs). 
At the later epoch (right panel), the polar region
of the gas disk is highly dissociated by the radiation from the sink
particles. The dissociated region is heated up to $7000-8000$K, 
due to the accretion shocks and the chemical heating.
In the absence of photo-dissociating radiation,
H$_2$ dissociation cooling should almost cancels the H$_2$ formation heating.
But, in the photo-dissociated region, H$_2$ formation heating 
is of great significance, since H$_2$ dissociation cooling is negligible
in such environments.  

The ``heated bubble'' originating from 
the absence of coolant as well as the chemical heating by
photo-dissociation reduces the mass accretion rate onto the sink particles. 
In fact, the total accreted mass after 24000 yrs is $\sim 50\%$ smaller
than the mass without radiative feedback (Fig. \ref{fig:total_sink_mass}). Thus, the
photo-dissociation is one of the important effects to control the final
mass of the first stars. On the other hand, it is worth noting that the
photo-dissociation feedback cannot quench
the fragmentation of the gas disks. We observe the fragmentation of the disk
in both cases with and without radiative feedback, and 
the numbers of the sink particles are $\sim 10$.

We also find that a few sinks are kicked away from the central part of
the host gas cloud via the gravitational three-body interaction. As a
result, the mass accretion onto these sinks is basically shut off. We find
that some of these sinks do not grow to more than 0.8$M_\odot$. 
The lifetime of main sequence of such low mass stars is $\sim 2 \times 10^{10}$yr. 
If we regard these sink particles as ``stars'', we might be able to find the
``real first stars'' in the Galactic halo.

\section{Evolution of First Galaxies by Internal UV feedback}

UV radiation affects the chemo-thermal and dynamical evolution of galaxies 
through the photo-ionization, -heating, and -dissociation processes. 
Although the importance of such UV feedback on structure formation 
in the universe have been pointed out already in 1980's, 
no calculation of three-dimensional hydrodynamics coupled with
radiative transfer has ever been realized until 2000's. 
In the last decade, it became possible to perform 
three-dimensional radiation hydrodynamics (3D-RHD) simulations, thanks to the developments 
of computers and calculation algorithms \cite{rf:Iliev09}. 
The {\it START} scheme enables us to solve 3D-RHD 
for numerous radiation sources \cite{rf:HU10}. 
In this section, 
we present the results of RHD simulations on the ionization of first galaxies 
and the escape of ionizing photons, and thereby demonstrate 
the necessity of RHD simulations.
Here, a $\Lambda$CDM cosmology with
$(\Omega_\Lambda, \Omega_{\rm m}, \Omega_{\rm b}, h)
=(0.73, 0.27, 0.049, 0.71)$ is employed.

\subsection{Determination of radiative reaction rates}

Once optical depths are determined, radiative reaction rates
such as the photo-ionization and photo-heating rates can be evaluated. 
For instance, the ionization rate for $i$-th chemical species 
(i.e., HI, HeI, or HeII) at a position $\bold r$ is generally given by 
\begin{equation}
\label{eq1}
	k_{ion, i}(\bold r) = n_{i}({\bold r})\int_{\nu_{L,i}}^{\infty}\int 
\sigma_{\nu, i}\frac{I_{\nu,0}e^{-\tau_{\nu}({\bold r})}}{h \nu}
	d\nu d\Omega,
\end{equation}
where $n_i$, $\nu_{L,i}$, and $\sigma_{\nu, i}$ are the number density, 
the Lyman limit frequency, and the cross-section for $i$-th species. 
$I_{\nu,0}$ is the intensity of incident radiation at frequency $\nu$, and
$\tau_{\nu}$ is the optical depth.
Here, we should carefully treat this equation, because it becomes numerically 
zero where $\tau_{\rm \nu}(\bold r)$ is much larger than unity. 
It implies that the ionization rates become zero even if the number 
of ionizing photons is enough to ionize the medium. 
This leads to the unphysical feature that the ionization front does not
propagate into an optically thick media. 
To avoid this problem, the form given by 
\begin{equation}
\label{eq2}
	k_{ion, i}(\bold r) = -\frac{1}{4\pi r^2}\frac{d}{dr}\int_{\nu_{L, i}}
	^{\infty}\frac{n_{i}(\bold r)\sigma_{\nu, i}}{(n\sigma)_{tot}}
\frac{L_{\nu}e^{-\tau_{\nu}(\bold r)}}{h\nu}d\nu, 
\end{equation}
is often used instead of equation (\ref{eq1}) so as to conserve the photon number. 
The photo-heating rates can be similarly evaluated by multiplying the term 
in the integral by $(h\nu-h\nu_{L,i})$. 

\subsection{Chemo-Thermal Evolution coupled with RHD}

We update the information of radiation field at every period of
$t_{RT}=0.1\times {\rm min}(t_{\rm rec}, t_{\rm hydro})$ 
by solving radiative transfer (RT), where $t_{\rm rec}$ and $t_{\rm hydro}$
are respectively the timescales for the recombination $t_{\rm rec}=1/n\alpha(T)$ 
and hydrodynamics $t_{\rm hydro}=h/(| {\bold v} |+c_{s}(T))$, 
where $n$, $T$, $h$, $\bold v$, and $c_{s}(T)$ are the number density, temperature, 
smoothing length, velocity, and sound velocity of a SPH particle, respectively. 
In a case of $t_{\rm hydro} > t_{\rm rec}$, we repeatedly solve the RT 
at an interval of $t_{\rm rec}$ during $t_{\rm hydro}$.  
In order to determine $t_{\rm rec}$,
we always use only particles on which the relative change of ionized fraction 
during previous $t_{RT}$ is greater than 10 percent. 
This treatment allows us to effectively reduce the computational cost due to the following two reasons. 
First, it is often true that high density regions are well self-shielded, 
thus $t_{\rm rec}$ should be determined among SPH particles 
that reside outside of the self-shielded regions, otherwise we have to solve RT many times to no avail. 
Secondly, at a later phase of expansion of an HII region, which is called D-type ionization front, 
the ionization equilibrium is almost fulfilled in the HII region, and the change of radiation field is 
caused mainly by the hydrodynamic motion driven by the thermal pressure. 

When the UV feedback is included, the chemical timescale ($t_{\rm chem,i}=n_{i}/\dot{n_{i}}$) 
and thermal one ($t_{\rm thermal} = T/ \dot{T}$) are generally shorter than $t_{RT}$.  
In addition, the set of equations which determine the abundance of chemical species are stiff. 
Therefore, we implicitly solve the chemo-thermal evolution with subcycle 
timestep $t_{sub} = {\rm min}(t_{\rm chem},t_{\rm thermal})$. 
Here note that the ionization rates also vary during $t_{RT}$, since 
they are multiplied by $n_{i}=f_{i}n/(\mu m_{\rm H})$, where $f_i$ and $\mu$ 
are the fraction of $i$-th species and the mean molecular weight, respectively. 

\subsection{Feedback and escape of ionizing radiation in first galaxies}

It is important for revealing the cosmic reionization history to clarify 
how many ionizing photons can escape from galaxies. 
Previous numerical simulations that evaluate the escape fractions of 
ionizing photons from galaxies can be categorized into two types: 
One is RDH simulations in which UV feedback is consistently taken 
into account \cite{rf:Gnedin08,rf:WC09}, and the other is non-RHD simulations 
in which the transfer of ionizing photons is solved without the back-reactions 
on chemodynamics caused by UV feedback \cite{rf:RS10,rf:Yajima11}.  
The UV feedback is expected to play a crucial role on the determination 
of the escape fraction, since the gas in a less massive galaxy is 
easily evacuated by the UV feedback \cite{rf:WC09}. 
Thus it is generally expected that the escape fraction from a less massive 
galaxy is enhanced by the UV feedback, although nobody has quantitatively 
clarified how much the UV feedback impacts the escape fraction. 
On the other hand, the escape fraction from a massive galaxy is often expected 
to be insensitive to the UV feedback, since the galaxy hardly loses its gaseous components.  
However, if the internal density structure of the galaxy is significantly changed 
by the UV feedback, the escape fraction is possibly changed \cite{rf:FS11}. 

Here, we investigate the impact of UV feedback on the escape fraction of ionizing photons from a massive halo. 
We have performed a cosmological RHD simulation \cite{rf:HS12}, where the radiative transfer 
of UV radiation from individual stellar particles is solved together with the hydrodynamics.
In order to evaluate an ionizing photon production rate that depends on stellar age, 
we have calculated population synthesis with {${\rm P\acute{E}GASE}$} \cite{rf:FV97} 
in advance of the simulation, and used the obtained rate in the simulation. 
The masses of the SPH and dark matter particles are respectively 
$4.8\times 10^4M_{\odot}$ and $2.9\times 10^5M_{\odot}$.
Thus, we can resolve low mass halos down to $M_{\rm halo}\approx 3\times10^7M_{\odot}$  with 100 particles. 
For the reference case, we have also performed a pure hydrodynamic (non-RHD) simulation 
for the same initial conditions and cosmological parameters,
where the UV radiation transfer is not coupled and therefore no effect by UV feedback 
is included. In a non-RHD simulation, the escape fraction of ionizing photons is assessed
by solving the UV radiative transfer as a post process using the density distributions obtained by
a pure hydrodynamic simulation.

\begin{table}[b]
\begin{center}
\caption{Properties of simulated halos}
\begin{tabular}{lccccccc} \hline \hline
 Name & $M_{\rm halo}[M_{\odot}]$ &  $M_b [M_{\odot}]$ &  $M_{*} [M_{\odot}]$ & $\dot{N}_{int}[1/{\rm s}]$ & 
 $f_{\rm e, HI}^{a)}$ & $f_{\rm e, HeI}^{b)}$ & $f_{\rm e, HeII}^{c)}$ \\[1mm] \hline
 Halo-R (w/ UV)     &  $6.8\times 10^9 $ & $1.2 \times 10^9$ & $2.9 \times 10^7 $& $5.9\times  10^{52}$& 0.31   &  0.35 & 0.13  \\ 
 Halo-H (w/o UV)    &  $6.7\times 10^9$ & $1.0 \times 10^9$ & $1.0 \times 10^8$ &$3.6\times  10^{53}$& 0.15   &  0.14 & 0.11  \\ \hline
\label{table:rad1} 
\end{tabular}
\end{center}
{\scriptsize
$a)$ Escape fraction of photons ionizing HI. \\
$b)$ Escape fraction of photons ionizing HeI. \\
$c)$ Escape fraction of photons ionizing HeII. 
}
\end{table}

We choose the most massive halo at $z=6.0$ in each of the simulations. 
Hereafter, we call the halo in RHD simulation Halo-R, and that in the non-RHD simulation Halo-H. 
The properties of these halos are summarized in Table \ref{table:rad1}. 
Here, we should mention that the halos are sufficiently massive and 
hardly lose their baryonic components via photo-evaporation.
Actually,  the resultant ratio of the baryon mass to the halo mass is $M_{b}/M_{\rm halo}\approx 0.16$, 
which is comparable to $\Omega_{b}/\Omega_{\rm m}=0.18$.  

\begin{figure}[t]
\begin{center}
\includegraphics[width=13.5cm,]{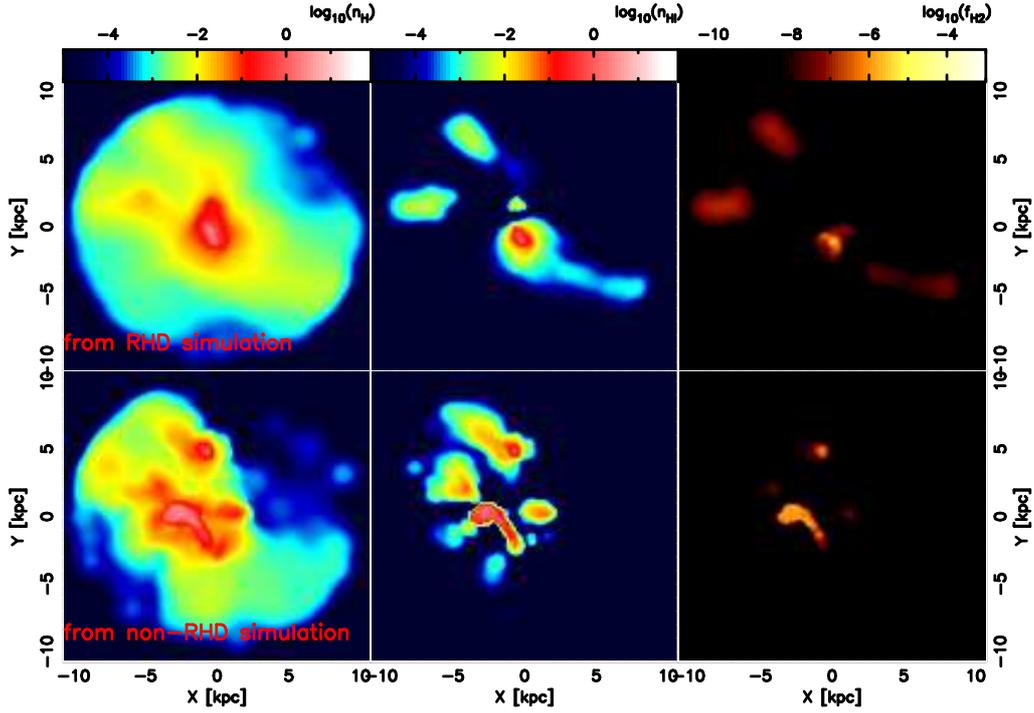}
   \caption{Maps of hydrogen number densities, HI number densities and $\rm H_2$ 
   fractions are shown from left to right. 
   The slice thickness is 2kpc.     
   The upper panels show the the halo taken from the RHD cosmological simulation (Halo-R), while 
   the  bottom panels do the halo taken from the non-RHD cosmological simulation (Halo-H) 
   with the post-processing RT calculation done by the procedure described in the text. }
\label{fig:rad1}
\end{center}
\end{figure}

Fig. \ref {fig:rad1} shows the numerical results, where number densities, ionized fractions, 
and $\rm H_2$ fractions are shown. 
For Halo-R, the ionized structures are calculated in the RHD simulation, and
the escape fractions of HI, HeI, and HeII ionizing photons are obtained. 
On the other hand, 
in order to assess the escape fractions of ionizing photons for Halo-H, 
we solve the radiative transfer as post-processing
until the ionization equilibrium is reached at all positions. 
Hereafter, we call this procedure the post-proccesing RT. 
This is a method frequently used to estimate the escape fraction\cite{rf:RS10,rf:Yajima11}. 
We count the number of escaping photons at $2\times64\times64$ virtual grid points placed outside the halo.
The upper panels in Fig. \ref {fig:rad1} show the RHD simulation, while 
lower panels are the non-RHD simulations with the post-processing RT calculation.

As seen in Fig. \ref {fig:rad1}, a high density peak forms near the center in Halo-R,
where the gas is self-shielded from ionizing radiation and H$_2$ molecules 
form efficiently. This central peak allows further star formation.
The gas in remaining regions is quite smoothly distributed and mostly ionized.
This is thought to be the consequence of radiation hydrodynamic feedback by
photo-ionization of H, photo-dissociation of H$_2$, and photo-heating. 
On the other hand, several density peaks appear in Halo-H,
where the gas in each peak is self-shielded from ionizing radiation. 
Recently, Fernandez \& Shull (2011) have pointed out that the escape fraction sensitively 
depends on the internal structure in a galaxy \cite{rf:FS11}.
The clumpy structure in the present simulation can reduce the escape fraction of ionizing photons from the galaxy.
The resultant escape fractions averaged over all directions are presented in Table \ref{table:rad1}.
The escape fractions from the Halo-R are actually higher than those from the Halo-H by a factor of 2. 
It has been often argued that UV feedback would increase the escape fractions 
for a less massive halo because of the mass loss by UV heating\cite{rf:WC09}.
Interestingly, the present results show that the escape fractions can be enhanced by UV feedback even in a massive halo. 
As for the reionization, what is important is the product of the intrinsic ionizing photon production rate $\dot{N}_{\rm int}$ 
multiplied by the escape fraction $f_{\rm e}$. 
The product is three times higher in the Halo-H, 
since the star formation rate is significantly higher in the Halo-H.

The dependence of UV feedback on the galaxy mass is unclear
in the present analysis, and hence a statistical study is needed. 
But, naively the impacts of UV feedback on the escape fractions 
and star formation are expected to be more dramatic for less massive haloes. 
Therefore, the present results demonstrate that RHD simulations are requisite to reveal the sources of the cosmic reionization. 


\section{Summary}

In this paper, we have explored the formation of first objects and first galaxies
with radiative feedbacks.
The mass of first objects has been investigated by high-resolution hydrodynamic simulations.
As a result, it has been revealed that the dark matter cusp potential, if it is resolved well, 
allows a smaller mass object to collapse, and then the first object mass could be 
reduced by roughly two orders of magnitude than in previous predictions.
This implies that the number of Pop III stars formed in the early universe 
could be significantly larger than hitherto thought.
Furthermore, radiation hydrodynamic simulations with the feedback by 
photo-ionization and photo-dissociation photons 
have shown that multiple stars can form in a $10^5M_\odot$ halo.
Also, the analytic criteria for radiation hydrodynamic feedback are presented.

Besides, the fragmentation of an accreting disk around a first protostar 
has been studied with photo-dissociation feedback. As a result, it is found that
the fragmentation is not quenched by photo-dissociation feedback,
but the ``heated bubble'' resulting from the photo-dissociation 
can reduce the mass accretion rate onto protostars. 
Also, protostars as small as 0.8$M_\odot$ can be ejected through
gravitational three-body interaction and evolve with keeping their mass.
Such small stars might be detected as ``real first stars'' in the Galactic halo.

Finally, radiation hydrodynamic simulations have been performed
to assess the impact of UV feedback on first galaxies. 
By comparing radiation hydrodynamic simulations  
to pure hydrodynamic simulations, we have found
that UV feedback enhances the escape fraction by a factor of 2 even in a massive halo.
But, the product of the intrinsic ionizing photon production rate  
multiplied by the escape fraction is three times higher in a case 
without UV feedback than in a case with UV feedback. 
These results imply that UV feedback deserves careful consideration
to reveal the cosmic reionization history.

\section*{Acknowledgements}

Numerical simulations have been performed with the {\it FIRST} simulator
and {\it T2K-Tsukuba} at Center for Computational Sciences in University of Tsukuba,
and also Blue Gene/P Babel at the SNRS computing center: IDRIS. 
This work was supported in part by the {\it FIRST} project based on
a Grants-in-Aid for Specially Promoted Research in 
MEXT (16002003) and a JSPS Grant-in-Aid for Scientific Research (S) (20224002),
by a JSPS Grant-in-Aid for Young Scientists B: 24740114, by MEXT HPCI STRATEGIC
PROGRAM, and by the French funding agency ANR (ANR-09-BLAN-0030).


\begin{thebibliography}{99}


\bibitem{Tegmark97}
M. Tegmark et al. \JL{Astrophysical J.,474,1997,1}

\bibitem{Fuller00}
T.M. Fuller and H.M.P. Couchman, \JL{Astrophysical J.,544,2000,6}

\bibitem{Yoshida03}
N. Yoshida, T. Abel, L. Hernquist, and N. Sugiyama, 
\JL{Astrophysical J.,592,2003,645}

\bibitem{SU06} 
H. Susa, H., and Umemura, M., \JL{Astrophysical J. Letters,645,2006,L93}

\bibitem{Hasegawa09} 
K. Hasegawa, M. Umemura, and H. Susa, 
\JL{Notices of the Royal Astronomical Society,395,2009,1280}

\bibitem{Susa09} 
H. Susa, M. Umemura, and K. Hasegawa, \JL{Astrophysical J.,702,2009,480} 

\bibitem{rf:Kitayama04}
T. Kitayama, N. Yoshida, H. Susa, M. Umemura, \JL{Astrophysical J.,613,2004,631}

\bibitem{rf:Yoshida07}
N. Yoshida, S. P. Oh., T. Kitayama, L. Hernquist, \JL{Astrophysical J.,663,2007,687}

\bibitem{ON98}
K. Omukai, and R. Nishi, \JL{Astrophys. J.,508,1998,141}

\bibitem{BCL99} 
V. Bromm, P. S. Coppi, and R. B. Larson,  \JL{Astrophysical J.,527,1999,L5}

\bibitem{Abel00} 
T. Abel, G. ~L. Bryan, and M.~L. Norman, \JL{Astrophysical J.,540,2000,39}

\bibitem{Abel02} 
T. Abel, G.~L.Bryan, and M.~L. Norman, \JL{Science,295,2002,93}

\bibitem{Yoshida08} 
N. Yoshida, K. Omukai, and L. Hernquist, \JL{Science,321,2008,669}

\bibitem{NU01} 
F. Nakamura, and M. Umemura, \ 2001, \JL{Astrophysical J.,548,2001,19}

\bibitem{ON07}
B.~W. O'Shea, and M.~L. Norman, \JL{Astrophysical J.,654,2007,66}

\bibitem{Hosokawa11}
T. Hosokawa, K. Omukai, N. Yoshida, and H. W. Yorke, \JL{Science,334,2011,1250} 

\bibitem{Schaerer02}
D. Schaerer, \JL{Astronomy \& Astrophysics,382,2002,28}

\bibitem{Clark11} 
P.~C. Clark, S.~C.~O. Glover, R.~J. Smith, T.~H. Greif, R.~S. Klessen, 
and V. Bromm, \JL{Science,331,2011,1040}

\bibitem{Stacy10}
A. Stacy, T.~H. Greif, and V. Bromm, 
\JL{Monthly Notices of the Royal Astronomical Society,403,2010,45} 

\bibitem{Greif11}
T.~H. Greif, V. Springel, S.~D.~M. White, et al. \JL{Astrophysical J.,737,2011,75} 

\bibitem{FMK05}
T. Fukushige, J. Makino, and A. Kawai, \JL{Publications of the Astronomical Society of Japan,57,2005,1009} 

\bibitem{Umemura08b} 
M. Umemura, H. Susa, T. Suwa, and D. Sato, \JL{AIP Conference Proceedings,990,2008,386}

\bibitem{Umemura08a}
M. Umemura et al. Proc. of IAU Symp. 255, Cambridge University Press
(on-line version) (2008).

\bibitem{Umemura10} 
M. Umemura, H. Susa, and T. Suwa, \JL{AIP Conference Proceedings,1238,2010,101}

\bibitem{NFW}
J. F. Navarro, C.S. Frenk, and S.D.M. White, \JL{Astrophysical J.,490,1997,493} 

\bibitem{FM97}
T. Fukushige, and J. Makino, \JL{Astrophys. J.,477,1997,L9} 

\bibitem{Moore}
B. Moore, S. Ghigna, F. Governato, G. Lake, T. Quinn, J. Stadel, and P. Tozzi, 
\JL{Astrophysical J.,524,1999,L19} 

\bibitem{JS00}
Y. P. Jing, and Y. Suto, \JL{Astrophys. J.,529,2000,69} 

\bibitem{FKM04}
T. Fukushige, A. Kawai, and J. Makino, \JL{Astrophys. J.,606,2004,625} 

\bibitem{Diemand08}
J. Diemand, et al.,  \JL{Nature,454,2008,735}

\bibitem{Springel08}
V. Springel, et al., \JL{Monthly Notices of the Royal Astronomical Society,391,2008,1685} 

\bibitem{DB96}
B. T. Draine, and F. Bertoldi, \JL{Astrophys. J.,468,1996,269} 

\bibitem{Kahn54} 
F.~D. Kahn, \JL{Bull. Astron. Inst. Netherlands,12,1954,187}

\bibitem{Susa07} 
H. Susa H., 2007, \JL{Astrophys. J.,659,2007,908}

\bibitem{Susa06}
H. Susa, \JL{Publications of the Astronomical Society of Japan,58,2006,445} 

\bibitem{bate_burket97}
M. Bate, and A. Burkert, \JL{Monthly Notices of the Royal Astronomical Society,288,1997,1060} 

\bibitem{rf:WH}
J. Wolcott-Green \& Z. Haiman, \JL{Monthly Notices of the Royal Astronomical Society,412,2011,2603} 

\bibitem{rf:HYO}
T. Hosokawa, H.W. Yorke, K. Omukai, \JL{Astrophys. J.,721,2010,478} 

\bibitem{rf:Iliev09}
I. Iliev, et al., \JL{Monthly Notices of the Royal Astronomical Society,400,2009,1283} 

\bibitem{rf:HS12}
K. Hasegawa, B. Semelin, Monthly Notices of the Royal Astronomical Society, submitted 

\bibitem{rf:HU10}
K. Hasegawa, M. Umemura, \JL{Monthly Notices of the Royal Astronomical Society,407,2010,2362} 

\bibitem{rf:KB00}
O. Kessel-Deynet, A. Burkert, \JL{Monthly Notices of the Royal Astronomical Society,315,2006,713} 

\bibitem{rf:BH86}
J. Barnes, P. Hut, \JL{Nature,324,1986,446}  

\bibitem{rf:Gnedin08}
N. Y. Gnedin, A. V. Kravtsov, H-W. Chen, \JL{Astrophys. J.,672,2008,765} 

\bibitem{rf:WC09}
J. H. Wise, R. Cen, \JL{Astrophys. J.,693,2009,984}    

\bibitem{rf:RS10}
A. O. Razoumov, J. Sommer-Larsen, \JL{Astrophys. J.,701,2010,1239} 

\bibitem{rf:Yajima11}
H. Yajima, J-H. Choi, K. Nagamine, \JL{Monthly Notices of the Royal Astronomical Society,412,2011,411} 

\bibitem{rf:FS11}
E. R. Fernandez, J. M. Shull, \JL{Astrophys. J.,731,2011,20}    

\bibitem{rf:FV97}
M. Fioc, B. Rocca-Volmerange, \JL{A\&A, 326,1997,950}


\end{thebibliography}
\end{document}